\documentclass[doublecol]{epl2} 
\title{Thermo-Rotational Instability in Plasma Disks Around Compact Objects(Revision)}
\shorttitle{} %Insert here a short version of the title if it exceeds 70 characters

\author{B. Coppi}
\shortauthor{}

\institute{Massachusetts Institute of Technology, Cambridge, MA, USA\\}

\abstract{Differentially rotating plasma disks, around compact objects, that are imbedded in a ``seed''  magnetic field are shown to develop vertically localized ballooning modes that are driven by the combined radial gradient of the rotation frequency and vertical gradients of the plasma density and temperature.  When the electron mean free path is shorter than the disk height and the relevant thermal conductivity can be neglected, the vertical particle flows produced by of these modes have the effect to drive the density and temperature profiles toward the ``adiabatic condition'' where 
$\eta_{T}\equiv\left(dlnT/dz\right)/\left(dlnn/dz\right)=2/3$.  Here $T$ is the plasma temperature and $n$ the particle density.  The faster growth rates correspond to steeper temperature profiles $(\eta_{T}>2/3)$  such as those produced by an internal (e.g., viscous) heating process.  In the end, ballooning modes excited for various values of $\eta_{T}$ can lead to the evolution of the disk into a different current carrying configuration such as a sequence of plasma rings.}

\begin{document}
\maketitle
\section{I.  Introduction}
Axisymmetric plasma disks \cite{1,2} in the prevalent gravity of a central object that do not have internal currents and are imbedded in a ``seed'' vertical magnetic field can be described in their equilibrium state by one-fluid equations that are basically the same as those describing gaseous disks \cite{3}.  Then axisymmetric plasma ballooning modes that are contained vertically within the disk and are radially modulated over the ``Alfv\'{e}n scale distance'' $\left(\textsc{v}_{A}/\Omega\right)$  are identified as the most likely process leading the starting disk configuration to evolve into a current carrying one such as the axisymmetric sequence of rings found in Ref.[4].  The driving factors of the fastest modes, within this category of modes, are the combined effects of the vertical particle density and temperature gradients and the radial gradient of the rotation frequency.  The particle motion is assumed to be ``frozen'' to the magnetic field.  
\par
The relevant quasi linear theory shows that this kind of mode, that is referred to as the ``thermo-rotational instability'', produces a vertical particle inflow toward the equatorial plane and a temperature outflow if $\eta_{T}={\left(dlnT/dz\right)}/{\left(dlnn/dz\right)}>2/3$.  The reverse occurs if $\eta_{T}<2/3$ .  Here $T$ is the plasma temperature $\left(2T=p/n\right)$, $p$  is the total plasma pressure and $n$ is the particle density.  Therefore, the main effect of these modes is to drive the density and temperature profiles toward the ``adiabatic'' condition  $\eta_{T}=2/3$.  When this condition is reached the slowly growing ballooning modes presented in Ref. \cite{5} can take over and produce, at saturation, a new stationary axisymmetric configuration\cite{4}.
\par
In this context we note that the well known Magneto-Rotational Instability (M.R.I.), driven simply by the gradient of the rotation frequency $\Omega$, was derived originally for a cylindrical geometry \cite{6,7,8} where it can develop a growth rate close to $\Omega$ and produce a significant rate of radial transport of angular momentum.  On the other hand this instability is severely depressed when a strong bending of the magnetic field lines, imposed by the geometry of a thin disk, is considered.  When $\eta_{T}>2/3$ , corresponding to a temperature profile for a disk subject to a strong heating source (e.g. viscous) around the equatorial plane, the growth rate of the ballooning thermo-rotational instability is not too inferior to that of the ``cylindrical'' M.R.I. for equal value of the rotation frequency.  We note also that the thermo-rotational instability would not be found if the flow compressibility were neglected.

\section{II. Initial (Stationary) Configuration}

We consider a current-less plasma disk whose density profile near the equatorial plane at the reference radial distance  $R_0$ from the axis of symmetry is represented by
\begin{equation}
\label{eq.1} 
\rho\left(z^{2},R_{0}\right)\simeq\rho_{0}\left(1-z^{2}/H^{2}_{0}\right),
\end{equation}
where $\eta_{T}$ can be positive as well as nul and negative and $H^{2}_{0}\ll R^{2}_{0}$.  This approximation is consistent with the analysis of ballooning modes of the type introduced in Refs. \cite{9,5} that are localized over a height of the order of $\Delta_{z}$ such that $\Delta^{2}_{z}<H^{2}_{0}$. For the sake of simplicity, we assume that the disk is immersed in a vertical magnetic field and that no appreciable internal currents are present.  Then the initial state is simply described by the vertical and radial components of the total momentum conservation that, for a thin disk in which no poloidal velocities are present, reduce to 
\begin{equation}
\label{eq.2}
0\simeq -\frac{\partial p}{\partial z}-z\Omega^{2}_{k}\rho
\end{equation}
and 
\begin{equation}
\label{eq.3}
\Omega\left(R\right)=\Omega_{k},
\end{equation}
where $\textsc{v}_{\phi}=\Omega\left(R\right)R$ is the plasma toroidal velocity, $\Omega_{k}=\left(GM_{\ast}/R^{3}_{0}\right)^{1/2}$ is the Keplerian frequency, $M_{\ast}$ is the mass of the central object, $p=2nT$, $\rho=nm_{i}$ and $m_{i}$ is the ion mass.   In particular, Eq.(2) yields $H^{2}_{0}=2T_{0}\left(1+\eta_{T}\right)/\left(m_{i}\Omega^{2}_{k}\right)$.
\\
\par
We consider axisymmetric perturbations from the equilibrium state represented by 
\begin{equation}
\label{eq.4}
\hat\textsc{v}_{R}=\hat\textsc{v}_{R}\left(R-R_{0},z\right)\exp\left(\gamma_{0}t\right), 
\end{equation}
and we make use of the following total momentum conservation equation
\begin{eqnarray}
\label{eq.5}
\hat\mathbf{A}_{m}\equiv\rho\left(\frac{\partial\hat\mathbf{v}}{\partial t}+\mathbf{v}\cdot\nabla\hat\mathbf{v}+\hat{\mathbf{v}}\cdot\nabla\mathbf{v}\right)
+\nabla\left(\hat{p}+\frac{\hat\mathbf{B}\cdot\mathbf{B}}{4\pi}\right)\nonumber\\
\nonumber\\
-\frac{1}{4\pi}\mathbf{B}\cdot\nabla\hat\mathbf{B}+z\hat\rho\Omega^{2}_{k}\mathbf{e}_{z}=0.
\end{eqnarray}

We adopt the ``hyperconductivity'' condition
\begin{equation}
\label{eq.6}
\hat\mathbf{E}+\left(\hat\mathbf{v}\times\mathbf{B}+\mathbf{v}\times\hat\mathbf{B}\right)/c=0
\end{equation}
and write 
\begin{equation}
\label{eq.7}
\hat\mathbf{v}_{p}=\gamma_{0}\boldsymbol{\hat\xi}_{p}
\end{equation}
where $\hat\mathbf{v}_{p}$ is the poloidal velocity and $\boldsymbol{\hat\xi}_p$ the relevant displacement. Thus taking the curl of Eq.(6) we obtain
\begin{equation}
\label{eq.8}
\hat\textit{B}_{R} = B_{z}\frac{\partial}{\partial z}\hat{\xi}_{R},\hspace{1cm}\hat\textit{B}_{\phi} = B_{z}\frac{\partial}{\partial z}\hat{\xi}_{\phi}
\end{equation}
\begin{equation}
\label{eq.9}
\hat\textit{B}_{z}=\textit{B}_{z}\frac{\partial}{\partial  z}\hat\xi_{z}-{B}_{z}\left(\nabla\cdot\boldsymbol{\hat\xi}_{p}\right),
\end{equation}
where 
\begin{equation}
\label{eq.10}
\gamma_{0}\hat\xi_{\phi}=\hat\textsc{v}_{\phi}+R\frac{d \Omega}{d R}\hat\xi_{R},
\end{equation}
and $\hat\textsc{v}_{\phi}$  is the perturbed toroidal velocity.  Then we follow the same proceedure as that in Ref.\cite{5} considering high$-\beta$ regimes where $\textsc{v}^{2}_{A}<c^{2}_{s}$, $c_{s}$ is the sound velocity and $\textsc{v}_{A}=B/\left(4\pi\rho\right)^{1/2}$ is the Alfv\'{e}n velocity.   As pointed out in Ref. \cite{9,5}the present problem involves prominently the scale distance $\textsc{v}_{A}/\Omega_{k}$. 

\section{III. Importance of Thermal Fluctuations}

The z-component of the total momentum conservation equation can be written as
\begin{equation}
\label{eq.11}
\gamma^{2}_{0}\rho\hat{\xi}_{z}=-\frac{\partial \hat p}{\partial z}-z\hat{\rho}\Omega^{2}_{k}
\end{equation}
and we note that the two terms on the r.h.s. scale as  
\begin{equation}
\label{eq.12}
\hat p/\left(p\Delta_{z}\right)\phantom{i}\textnormal{and}\phantom{ii}\hat\rho\Delta_{z}/\left(\rho H^{2}_{0}\right),
\end{equation}	
relative to each other.  Therefore, if we consider cases, as was done in Refs.\cite{5}, where
\begin{equation}
\label{eq.13}
\left|\frac{\hat{p}}{p}\right|\sim\left|\frac{\hat\rho}{\rho}\right| \geq\left|\frac{\hat{T}}{T}\right|,
\end{equation}
the term $\left|\hat\rho g_{z}\right|$, where $g_{z}\equiv-z\Omega^{2}_{k}$, can be neglected rela{\-}tively to$\left|\partial \hat{p}/\partial z\right|$ . 
\\
\par
On the other hand, if $\left|\hat\rho/\rho\right|\sim\left(H^{2}_{0}/\Delta^{2}_{z}\right)\left|\hat{p}/p\right|$ implying that 
\begin{equation}
\label{eq.14}
\hat\rho/\rho\simeq -\hat{T}/T, 
\end{equation}
the $\hat\rho g_{z}$ term has to be retained. Since $\hat{p}/p=\hat\rho/\rho+\hat{T}/T$,
\begin{equation}
\label{eq.15}
-\frac{\partial}{\partial z}\hat{p}-z\Omega^{2}_{k}\hat\rho\simeq-\frac{\partial}{\partial z}\hat{p}+z\Omega^{2}_{k}\rho\frac{\hat{T}}{T}.
\end{equation}
Thus the $g_{z}$ term remains associated with the temperature perturbations.
\\
\par
In this connection we note that a simplified thermal energy balance equation that includes an isotropic thermal conductivity is
\begin{eqnarray}
\label{eq.16}
\left(\gamma_{0}+D_{th}k^{2}_{R}\right)\hat{T}+\gamma_{0}\hat{\xi}_{z}\frac{dT}{dz}\nonumber\\
\nonumber\\
-\frac{2}{3}\gamma_{0}\left[\frac{\hat{p}}{n}-{\hat{T}}+\hat{\xi}_{z}\frac{d \rho}{dz}\frac{T}{\rho}\right]\simeq 0
\end{eqnarray}
where $D_{th}$ is the relevant thermal diffusion coefficient, considering perturbations that are appropriate for the modes we analyze, as shown in Ref. \cite{5} are locally periodic in $\left(R-R_{0}\right)$ and represented by
\begin{equation}
\label{eq.17}
\hat{T}\simeq\tilde{T}\left(z\right)\exp\left[\gamma_{0}t+ik_{R}\left(R-R_{0}\right)\right] ,
\end{equation}
where $k^{2}_{R}R^{2}_{0}\gg1$ and $\tilde{T}\left(z\right)$ is localized over $\Delta_{z}$ .  Then
\begin{eqnarray}
\label{eq.18}
-\frac{\partial}{\partial z}\hat{p}-\hat{\rho}\Omega^{2}_{k}z\simeq\frac{\partial}{\partial z}\hat{p}-z\Omega^{2}_{k}\rho\hat{\xi_{z}}\nonumber\\
\nonumber\\
\times\frac{3\gamma_{0}}{5\gamma_{0}+3\nu_{th}}\frac{1}{\rho}\frac{d \rho}{dz}\left(\eta_{T}-\frac{2}{3}\right),
\end{eqnarray}
where $\nu_{th}\equiv D_{th}k^{2}_{R}$.  Given Eq. (11), it is clear from the present result that the analysis given in Ref.\cite{5} is valid for
\begin{equation}
\label{eq.19}
\gamma_{0}>\Omega^{2}_{k}\frac{3\Delta^{2}_{z}}{H^{2}_{0}\left(5\gamma_{0}+3\nu_{th}\right)}\left|\eta_{T}-\frac{2}{3}\right|,
\end{equation}
considering that the growth rate that is found in this case is of the order of $\gamma_{0}\sim\textsc{v}_{A}/H_{0}$.  Clearly the inequality (19) holds either for relatively large thermal conductivities $\left(\gamma_{0}/\nu_{th}\ll 1\right)$  or ``adiabatic profiles'' such that $\left|\eta_{T}-2/3\right|\ll 1$.\\
\section{IV. Ballooning Mode Equations}
We define
\begin{equation}
\label{eq. 20}
C_{T}\equiv\frac{6}{5}\left(\eta_{T}-\frac{2}{3}\right)
\end{equation}
and consider it to have significant values of either signs.   Clearly, flat temperature profiles correspond to $C_{T}\simeq-4/5$.  Conversely, a heating process localized around the equatorial plane can be envisioned to maintain relatively peaked temperature profiles corresponding to $C_{T}>0$. 
We refer to Eq. (5) and note that the displacement  $\hat{\xi}_{\phi}$  is related to $\hat{\xi}_{R}$ by the $\phi$ - component of the total momentum conservation equation that is
\begin{equation}
\label{eq.21}
\frac{\gamma_{0}}{\textsc{v}^{2}_{A}}\left(\gamma_{0}\hat{\xi}_{\phi}+2\Omega_{k}\hat{\xi}_{R}\right)\simeq\frac{\partial ^{2}}{\partial z^{2}}{\hat\xi}_{\phi}.
\end{equation}
We consider the limit where
\begin{equation}
\label{eq.22} 
\gamma^{2}_{0}< \textsc{v}^{2}_{A}/\Delta^{2}_{z}
\end{equation}
that will be justified a posteriori.  Therefore Eq.(21) reduces to  
\begin{equation}
\label{eq.23} 
2\gamma_{0}\Omega_{k}\tilde{\xi}_{R}\simeq\textsc{v}^{2}_{A}\frac{d^{2}}{dz^{2}}\tilde{\xi}_{\phi}.
\end{equation}
The $\partial\left(\mathbf{e}_{\phi}\cdot\nabla\times\hat\mathbf{A}_{m}\right)/\partial z=0$ equation derived from the total momentum conservation equation $\hat\textbf{A}_{m}=0$ can be written as
\begin{eqnarray}
\label{eq.24}
\frac{\partial^2}{\partial z^2}\left[\rho\left(\gamma^{2}_{0}\hat{\xi}_{R}-2\Omega_{k}\hat{\textsc{v}}_{\phi}\right)
-\frac{B^{2}_{z}}{4\pi}\frac{\partial^{2}}{\partial z^{2}}\hat{\xi}_{R}\right]\nonumber\\
\nonumber\\
-\frac{\partial}{\partial z}\left\{\frac{\partial}{\partial R}\left[\rho\gamma^{2}_{0}\hat{\xi}_{z}
+z\Omega^{2}_{k}\hat{p}-\frac{B^{2}_{z}}{4\pi}\frac{\partial^2}{\partial R \partial z}\hat{\xi}_{R}\right]\right\}=0 .
\end{eqnarray}
We reconsider Eq. (14) that leads to
\begin{equation}
\label{eq.25} 
\nabla\cdot\boldsymbol{\hat{\xi}}\simeq-\left(3/5\right)\hat\xi_{z}\left(dp/dz\right)/p\sim\left|\hat\xi_{z}\right|\Delta_{z}/H^{2}_{0}
\end{equation}
and, consequently, to
\begin{equation}
\label{eq.26}
\frac{\partial}{\partial z}\hat{\xi}_{z}\simeq-\frac{\partial}{\partial R}\hat{\xi}_{R}.
\end{equation}
Then, making use of Eq. (23) we obtain
\begin{eqnarray}
\label{eq.27}
\frac{d^{2}}{dz^{2}}\left\{\rho\left(\gamma^{2}_{0}\tilde{\xi}_{R}+2\Omega_{k}\Omega^{\prime}_{k} R_{0}\tilde{\xi}_{R}\right)-\frac{B^{2}_{z}}{4\pi}\left(\frac{d^{2}}{dz^{2}}-k^{2}_{R}\right)\tilde{\xi}_{R}\right\}\nonumber\\
\nonumber\\
-\rho\gamma^{2}_{0}\left(k^{2}_{R}+\frac{4\Omega^{2}_{k}}{\textsc{v}^2_A}\right)\tilde{\xi}_{R}\simeq ik_{R}\Omega^{2}_{k}\frac{d}{dz}\left(z\tilde{\rho}\right),
\end{eqnarray}
where
\begin{equation}
\label{eq.28} 
\frac{\tilde\rho}{\rho}\simeq-z\tilde\xi_{z}\frac{C_{T}}{H^{2}_{0}}.
\end{equation}
Therefore, the r.h.s. of Eq. (27) becomes
\begin{equation}
\label{eq.29}
-k^{2}_{R}\Omega^{2} \rho \frac{C_{T}}{H^{2}_{0}}\left[z^{2}\tilde{\xi}_{R}+2z\left(\frac{i\tilde{\xi}_{z}}{k_{R}}\right)\right],
\end{equation}
where
\begin{equation}
\label{eq.30}
\frac{d}{dz}\left(i\frac{\tilde\xi_{z}}{k_{R}}\right)\simeq\tilde{\xi}_{R}.
\end{equation}
Now we define $\tilde{A}_{z}\equiv i\hat{\xi}_{z}/k_{R}$ considering the case where $k^{2}_{R}\texttt{v}^{2}_{A}\simeq-2\Omega_{k}\Omega^{\prime}_{k}R_{0}-2k_{0}$
$\left|\delta k_{R}\right|\textsc{v}^{2}_{A}$, $k_{0}>0$, for  $k^{2}_{0}\equiv-2\Omega_{k}\Omega^{\prime}_{k}R_{0}/\textsc{v}^{2}_{A}$, and localized modes for which $\rho$ can be represented by Eq. (1), Eq. (27) reduces to
\begin{eqnarray}
\label{eq.31}
\frac{d^{2}}{dz^{2}}\left\{\left[3\Omega^{2}_{k}\frac{z^{2}}{H^{2}_{0}}-\left(2k_{0}\left|\delta k_{R}\right|\textsc{v}^{2}_{A}\right)\right]\tilde{\xi}_{R}-\textsc{v}^{2}_{A}\frac{d^{2}}{dz^{2}}\tilde{\xi}_{R}\right\}\nonumber\\
\nonumber\\
-k^{2}_{0}\left(\frac{7}{3}\gamma^{2}_{0}\tilde{\xi}_{R}\right)
=-k^{2}_{0}\frac{\Omega^{2}_{k}}{H^{2}_{0}}C_{T}\frac{d}{dz}\left(z^{2}\tilde{A}_{z}\right),
\end{eqnarray}
as $-2R_{0}\Omega^{\prime}_{k}\Omega_{k}=3\Omega^{2}_{k}$.  By using Eq. (30) this can be rewritten as a quintic equation for $\tilde{A}_{z}$ that can be integrated to give the quartic
\begin{eqnarray}
\label{eq.32}
\frac{d}{dz}\left\{\left(\epsilon_{k}-\frac{z^{2}}{H^{2}_{0}}\right)\frac{d}{dz}\tilde{A}_{z}+\frac{1}{k^{2}_{0}}\frac{d^{3}}{dz^{3}}\tilde{A}_{z}\right\}\nonumber\\
\nonumber\\
+\left(\frac{7}{3}\frac{\gamma^{2}_{0}}{\textsc{v}^{2}_{A}}-C_{T}\frac{k^{2}_{0}z^{2}}{3H^{2}_{0}}\right)\tilde{A}_{z}=0,
\end{eqnarray}
where $\epsilon_{k}\equiv 2\left|\delta k_{R}\right|/k_{0}$.  The related quadratic form that can be used to evaluate $\gamma^{2}_{0}$ is
\begin{eqnarray}
\label{eq.33}
\frac{7}{3}\frac{\gamma^{2}_{0}}{\textsc{v}^{2}_{A}}\left\langle\left|\tilde{A}_{z}\right|^{2}\right\rangle=\frac{C_{T}k^{2}_{0}}{3H^{2}_{0}}\left\langle z^{2}\left|\tilde{A}_{z}\right|^{2}\right\rangle\nonumber\\
\nonumber\\
+\left\langle\left|\frac{d}{dz}\tilde{A}_{z}\right|^{2}\left(\epsilon_{k}-\frac{z^{2}}{H^{2}_{0}}\right)\right\rangle\ - \frac{1}{k^{2}_{0}}\left\langle \left|\frac{d^{2}}{dz^{2}}\tilde{A}_{z}\right|^{2}\right\rangle,
\end{eqnarray}
where $\left\langle\right\rangle\equiv\mathbf{\int}^{{H}_{0}}_{-{H}_{0}}dz$, we have considered modes that are localized over $\Delta_{z}<H_{0}$ and carried out appropriate integrations by part taking into account that $\tilde{A}_{z}^{*}d\tilde{A}_{z}/d z \rightarrow 0$ and $\tilde{A}_{z}^{*}d^{3}\tilde{A}_{z}/d z \rightarrow 0$ as $z^{2}\rightarrow H^{2}_{0}$.  Clearly, $\gamma^{2}_{0}$ is always real, that is, either purely growing or stable oscillatory modes can be found in the context of the adopted approximations.
\par
The relevant expression for $\tilde{p}\left(z\right)$ is obtained from the equation 
\begin{equation}
\label{eq.34}
\rho_{0}\left(\gamma^{2}_{0}+\frac{2z^{2}C_{T}}{H^{2}_{0}}\Omega^{2}_{k}\right)\tilde{A}_{z}\simeq\left(-\frac{d}{dz}\tilde{p}\right)\frac{i}{k_{0}},
\end{equation}
derived from Eq. (11) that gives $\tilde{p}$ as a function of $\tilde{\xi}_{z}$.
\section{V.	Fourier Transform}
The order of the equation to be solved can be reduced further, as we look for vertically localized modes, by taking the Fourier transform of Eq. (32).  For this, we adopt the following dimensionless variables
\begin{equation}
\label{eq.35}
\bar{z}\equiv\frac{z}{\Delta_{z}},\phantom{ii}
\Gamma^{2}_{0}=\frac{7}{9}\frac{\gamma^{2}_{0}}{\Omega^{2}}\left(k_{0}\Delta_{z}\right)^{4},\phantom{ii}
E^{0}_{k}\equiv \epsilon_{k}\left(k_{0}\Delta_{z}\right)^{2},
\end{equation}
where $\left(k_{0}\Delta_{z}\right)^{2}\gg1$, and rewrite Eq. (32) as
\begin{eqnarray}
\label{eq.36}
\frac{d}{d\bar{z}}\left\{\left(E^{0}_{k}-\bar{z}^{2}\frac{k^{2}_{0}\Delta^{4}_{z}}{H^{2}_{0}}\right)\frac{d}{d\bar{z}}\tilde{A_{z}}+\frac{d^{3}}{d\bar{z}^{3}}\tilde{A_{z}}\right\}\nonumber\\
\nonumber\\
+\left[\Gamma^{2}_{0}-C_{T}\left(\frac{k^{4}_{0}\Delta^{6}_{z}}{3H^{2}_{0}}\right)\bar{z}^{2}\right]\tilde{A_{z}}=0.
\end{eqnarray}
Then we define
\begin{equation}
\label{eq.37}
\Delta^{2}_{0}\equiv\frac{H_{0}}{k_{0}}\phantom{ii}\textnormal{and}\phantom{ii}C^{0}_{T}\equiv{C}_{T}\frac{\left(k_{0}\Delta_{z}\right)^{2}}{3}\left(\frac{\Delta_{z}}{\Delta_{0}}\right)^{4}
\end{equation}
and obtain the following equation for the Fourier transform $\tilde{A}_{k}$ of $\tilde{A}_{z}$ 
\begin{eqnarray}
\label{eq.38}
C^{0}_{T}\frac{d^{2}}{d\bar{k}^{2}}\tilde{A}_{k}-\left(\frac{\Delta_{z}}{\Delta_{0}}\right)^{4}\bar{k}\frac{d^{2}}{d\bar{k}^{2}}\left(\bar{k}\tilde{A}_{k}\right)\nonumber\\
\nonumber\\
+\left[\Gamma^{2}_{0}-\bar{k}^{2}\left(E^{0}_{k}-\bar{k}^{2}\right)\right]\tilde{A}_{k}=0.
\end{eqnarray}
The relevant quadratic form
\begin{eqnarray}
\label{eq.39}
\Gamma^{2}_{0}\left\langle \left|\tilde{A}_{k}\right|^{2}\right\rangle=C^{0}_{T}\left\langle \left|\frac{d}{d\bar{k}}\tilde{A}_{k}\right|^{2}\right\rangle+E^{0}_{k}\left\langle \left|\bar{k}\tilde{A}_{k}\right|^{2}\right\rangle\nonumber\\
\nonumber\\
-\left[\left\langle\bar{k}^{2}\left|k\tilde{A}_{k}\right|^{2}\right\rangle+\left(\frac{\Delta_{z}}{\Delta_{0}}\right)^{4}\left\langle \left|\frac{d}{d\bar{k}}\left(\bar{k}\tilde{A}_{k}\right)\right|^{2}\right\rangle\right]
\end{eqnarray}
may be more convienient than Eq. (33) in order to evaluate $\Gamma^{2}_{0}$, for given values of $E^{0}_{k}$ and $C^{0}_{T}$, using appropriate trial functions.  Here $\left\langle \right\rangle\equiv\boldsymbol{\int}^{\infty}_{-\infty}d\bar{k}$, and in deriving Eq. (39) we have considered that, for the localized solutions of Eq. (38), $\tilde{A}_{k}d\tilde{A}_{k}/d\bar{k}$ and $\bar{k}\tilde{A}_{k}d\left(\bar{k}\tilde{A}_{k}\right)/d\bar{k}\rightarrow 0$ as $\bar{k}^{2}\rightarrow\infty$.
\par
Then, if we refer to cases for which $\left|C_{T}\right|\sim1$ we may choose $\Delta^{6}_{z}=3\Delta^{4}_{z}/k^{2}_{0}$.  Consequently, $\left(\Delta_{z}/\Delta_{0}\right)^{4}=\left(3\epsilon_{0}\right)^{2/3}$, where $\epsilon_{0}\equiv1/\left(k_{0}H_{0}\right)<1$.  In particular $\Delta_{z}=\left(H_{0}/k^{2}_{0}\right)^{1/3}3^{1/6}$ and Eq. (38) can be re-written as 
\begin{eqnarray}
\label{eq.40} 
C^{0}_{T}\frac{d^{2}}{d\bar{k}^{2}}\tilde{A}_{k}-\left(3\epsilon_{0}\right)^{2/3}\bar{k}\frac{d^{2}}{d\bar{k}^{2}}\left(\bar{k}\tilde{A}_{k}\right)\nonumber\\
\nonumber\\
+\left[\left(\Gamma^{2}_{0}-E^{0}_{k}\bar{k}^{2}\right)+\bar{k}^{4}\right]\tilde{A}_{k}=0 .
\end{eqnarray}
If instead, we consider cases where $\left|C_{T}\right|<1$ we may choose $\Delta_{z}=\Delta_{0}$.  Then $C^{0}_{T}=C_{T}/\left(3\epsilon_{0}\right)$ and Eq. (38) becomes 
\begin{eqnarray}
\label{eq.41}
C^{0}_{T}\frac{d^{2}\tilde{A}_{k}}{d\bar{k}^{2}}-\bar{k}\frac{d^{2}}{d\bar{k}^{2}}\left(\bar{k}\tilde{A}_{k}\right)\nonumber\\
\nonumber\\
+\left[\Gamma^{2}_{0}-\bar{k}^{2}\left(E^{0}_{k}-\bar{k}^{2}\right)\right]\tilde{A}_{k}=0.
\end{eqnarray}
\par
The cases where $\left|C^{0}_{T}\right|\sim1$ correspond to the ``ordering of maximum information'' where all the terms in Eq. (41) are of the same order.  Consequently, these are the cases that we shall analyze most closely.   We note also that
\begin{equation}
\label{eq.42}
\tilde{\xi}^{k}_{R}=i\bar{k}\tilde{A}_{k}
\end{equation}
and that, referring to Eq. (21),
\begin{equation}
\label{eq.43}
\left(\gamma^{2}_{0}+\bar{k}^{2}\frac{\textsc{v}^{2}_{A}}{\Delta^{2}_{z}}\right)\tilde{\xi}^{k}_{\phi}=-i2\Omega_{k}\gamma_{0}\bar{k}\tilde{A}_{k} .
\end{equation}
Therefore, when we look for even solutions of Eq. (38) with $\tilde{A}_{k}=\tilde{A}^{0}_{k}\neq0$ for $\bar{k}=0$ it is important to include the term $\gamma^{2}_{0}\tilde{\xi}^{k}_{\phi}$ in Eq. (43) in order to avoid a singularity for $\tilde{\xi}^{k}_{\phi}$ at $\bar{k}=0$.  This means that for these cases $\tilde{\xi}_{\phi}$ will be localized over scale distances of the order of $\textsc{v}_{A}/\gamma_{0}$ and when these become of the order of $H_{0}$, the assumed approximation (1) for $\rho$ is no longer valid.

\section{VI.	Peaked Temperature Profiles}

At first we analyze the case where the temperature profile is relatively peaked and corresponds to $\eta_{T}>2/3$.  We refer to Eq. (41) and consider the lowest order solution
\begin{equation}
\label{eq.44}
\tilde{A}_{k}=\tilde{A}^{0}_{k}\exp\left(-\bar{k}^{2}/2\right).
\end{equation}
Then we have
\begin{equation}
\label{eq.45}
C^{0}_{T}\left(\bar{k}^{2}-1\right)-\bar{k}\left(\bar{k}^{3}-3\bar{k}\right)+\Gamma^{2}_{0}-\bar{k}^{2}E^{0}_{k}+\bar{k}^{4}=0
\end{equation}

which gives
\begin{equation}
\label{eq.46}
\Gamma^{2}_{0}\simeq C^{0}_{T}, \phantom{ii}\textnormal{i.e.,}\phantom{ii} \gamma^{2}_{0}\simeq\frac{\sqrt{3}}{7}\frac{\textsc{v}_A}{H_{0}}\frac{2}{5}\left(3\eta_{T}-2\right)\Omega
\end{equation}
and
\begin{equation}
\label{eq.47}
E^{0}_{k}=3+C^{0}_{T},\phantom{ii}\textnormal{i.e.,}\phantom{ii}\epsilon_{k}=3\epsilon_{0}+\frac{2}{5}\left(\eta_{T}-\frac{2}{3}\right).\end{equation}
Clearly, this mode is unstable only if $\eta_{T}>2/3$.  We note that, the marginally stable  $\left(\Gamma_{0}=0\right)$ modes that can be found for $C^{0}_{T}=0$ are given by the simple equation
\begin{equation}
\label{eq.48}
\frac{d^{2}}{d\bar{k}^{2}}\left(\bar{k}\tilde{A}_{k}\right)-\left(E^{0}_{k}-\bar{k}^{2}\right)\left(\bar{k}\tilde{A}_{k}\right)=0,
\end{equation}
which is the Fourier transform of that given originally in Ref. [9],
\begin{equation}
\label{eq.49}
\textsc{v}^{2}_{A}\frac{d^{2}\tilde{\xi}_{R}}{dz^{2}}-3\Omega^{2}_{k}\left(\frac{z^{2}}{H^{2}_{0}}-\epsilon_{k}\right)\tilde{\xi}_{R}=0.
\end{equation}
\par
Clearly, when $C^{0}_{T}<0$ the considered mode is purely oscillatory. When $\eta_{T}\sim1$ the growth rate that can be obtained from Eq. (41) for $C^{0}_{T}\gg1$ and $E^{0}_{k}\gg1$ can be considerable and not much smaller than that of the MRI instability in a cylindrical plasma \cite{6}.  We note also that, for this mode, $\tilde{\xi}^{k}_{\phi}$ as given by Eq. (43) requires that the term $\gamma^{2}_{0}$ be retained relative to $\bar{k}^{2}\textsc{v}^{2}_{A}/\Delta^{2}_{z}$ in order to avoid a singularity at $\bar{k}=0$.  This means that $\tilde{\xi}_{\phi}\left(z\right)$ is localized over a larger scale distance than $\Delta_{0}$, as indicated earlier.
\par
The higher eigenfunction
\begin{equation}
\label{eq.50}
\tilde{A}_{k}=\bar{k}\exp\left(-\bar{k}^{2}/2\right)
\end{equation}
remains unstable even for $C^{0}_{T}<0$ and corresponds to the first unstable mode analyzed in Ref. [5] for $C^{0}_{T}=0$.  In this case we obtain $\Gamma^{2}_{0}=2+3C^{0}_{T}$, i.e.
\begin{equation}
\label{eq.51}
\gamma^{2}_{0}\simeq\frac{6}{7}\frac{\textsc{v}_{A}}{H_{0}}\left[\frac{\sqrt{3}}{5}\left(3\eta_{T}-2\right)\Omega_{k}+\frac{\textsc{v}_{A}}{H_{0}}\right]
\end{equation}\\
and 
\begin{equation}
\label{eq.52}
E^{0}_{k}=5+C^{T}_{0}.
\end{equation}
\par
We observe that the relevant growth rate vanishes for $C^{T}_{0}<-2/3$, that is for relatively flat temperature profiles.  Moreover, in this case the $\gamma^{2}_{0}\tilde{\xi}^{k}_{\phi}$ contribution to Eq. (43) is not needed to avoid a singularity at $\bar{k}=0$ and $\tilde{\xi}^{k}_{\phi}\propto \exp\left(-\bar{k}^{2}/2\right)$.

\section{VII. 	Flatter Temperature Profiles}
When $\eta_{T}<2/3$ and $C^{T}_{0}$ is negative the higher even modes have to be considered.  For this we define $C^{0}_{0}=-C^{T}_{0}$ and we note that the first relevant solution in this class has the expression
\begin{equation}
\label{eq.53}
\tilde{A}_{k}=\tilde{A}^{0}_{k}\left(1+\alpha_{k}\bar{k}^{2}\right)\exp\left(-\bar{k}^{2}/2\right),
\end{equation}
where a necessary condition, in order to find $\Gamma^{2}_{0}>0$, is that $\alpha_{k}>1/2$ as can be seen from the asymptotic solution of Eq. (41) for $\bar{k}^{2}\rightarrow 0$.
\par
In particular, by inserting the expression (53) in Eq. (41) we obtain
\begin{equation}
\label{eq.54}
\Gamma^{2}_{0}=C^{0}_{0}\left(2\alpha_{k}-1\right),
\end{equation}
and
\begin{equation}
\label{eq.55}
E^{0}_{k}=7-C^{0}_{0}=3-C^{0}_{0}+\alpha_{k}\left(\Gamma^{2}_{0}+5C^{0}_{0}-6\right) .
\end{equation}
Thus
\begin{equation}
\label{eq.56}
\alpha_{k}=\frac{1}{2}\left(\frac{\Gamma^{2}_{0}}{C^{0}_{0}}+1\right)=\frac{4}{\left(\Gamma^{2}_{0}+5C^{0}_{0}-6\right)}
\end{equation}
and we obtain the dispersion relation
\begin{equation}
\label{eq.57}
\Gamma^{4}_{0}-6\left(1-C^{0}_{0}\right)\Gamma^{2}_{0}-C^{0}_{0}\left(14-5C^{0}_{0}\right)=0
\end{equation}
that gives
\begin{equation}
\label{eq.58}
\Gamma^{2}_{0}=3\left(1-C^{0}_{0}\right)+\sqrt{9\left(1-C^{0}_{0}\right)^{2}+C^{0}_{0}\left(14-5C^{0}_{0}\right)} .
\end{equation}
From this we see that for $C^{0}_{0}\rightarrow 0$, $\Gamma^{2}_{0}\rightarrow 6$ and that $\Gamma^{2}_{0}=0$ for $C^{0}_{0}=14/5$, that is $\Gamma_{0}$ decreases monotonically until it vanishes as $C^{0}_{0}$ increases from zero to $14/5$.  When $C^{0}_{T}<-14/5$ a higher eigensolution of Eq. (41) has to be considered. We note that, when $C^{0}_{0}\rightarrow0$, $\alpha_{k}\rightarrow 3/C^{0}_{0}$ while the corresponding expression for $\tilde{A}_{z}\left(\bar{z}\right)$ remains a well behaved function.
\par
Moreover, we can verify that if $C^{0}_{T}>0$ this mode remains unstable and that $\Gamma^{2}_{0}$ increases as $C^{0}_{T}$ increases.  

\section{VIII.	Vertical Transport}
The considered mode can produce a significant rate of particle density transport, in the vertical direction, that is of contrary sign to that of the temperature transport and modify the density and temperature profiles in such a way as to lead $\eta_{T}$ toward $2/3$.  Thus, if $\eta_{T}>2/3$ a particle inflow toward the equatorial plane is induced while the temperature rises away from the equatorial plane.  In fact, this is somewhat similar to the theoretical explanation\cite{10} for the particle inflow observed in magnetically confined toroidal plasmas that is associated with the radial electron temperature gradient.  In the present case we may argue that $\eta_{T}$ can be kept above $2/3$  by a  plasma heating process\cite{3}, such as one due to an anomalous viscosity, that is localized around the equatorial plane and attracts particles toward the equatorial plane.  A significant cooling process at the edge of the plasma disk can reinforce the described type of transport.
\par
When $\eta_{T}<2/3$, including the case where $\eta_{T}=0$ or where the surface of the disk can be hotter than the interior, a rise of the central temperature is induced while the particle transport is outward, from the equatorial plane.  These arguments are based on the relevant quasi-linear analysis that gives the vertical particle flux
\begin{eqnarray}
\label{eq.59}
\Gamma_{pz}=\left\langle\left\langle \hat{n}{\textsc{v}}_{z}\right\rangle\right\rangle\simeq-\frac{4}{5}\gamma_{0}\left\langle\left\langle \left|\hat{\xi_{z}}\right|^{2}\right\rangle\right\rangle\nonumber\\
\nonumber\\
\times\left[\frac{\partial}{\partial z}n-\frac{3}{2}\frac{n}{ T}\frac{\partial}{\partial z} T\right],
\end{eqnarray}
where $\left\langle \left\langle \right\rangle\right\rangle$ indicates an average over a radial distance $\Delta R$ such that $1/k_{0}<\Delta R<R_{0}$.  The corresponding temperature flux is $\left\langle \left\langle \hat{T}\hat\textsc{v}_{z}\right\rangle\right\rangle\simeq -\left\langle \left\langle \hat{n}\hat\textsc{v}_{z}\right\rangle\right\rangle T/n$.

\section{IX.	Conclusions}
The vertical density and temperature gradients in a differentially rotating plasma disk combined with those of the radial gradient of the rotation frequency are the key factors involved in the excitation of localized (ballooning) modes that are contained within the disk when a seed vertical magnetic field is present. The most important effect of these modes is that of inducing a change of both the density and the temperature profiles around the equatorial plane until these reach the ``adiabatic condition'' $(d{ln}T/dz/(d{ln}n/dz)=2/3$ .  Thus transport in the vertical direction is shown to have a relevant role in the dynamics of disks while, previously, only the radial fluxes of angular momentum and particles had been paid attention to. 
We note that the class of modes considered here benefits from the relatively strong vertical gradients of particle density and temperature that can be present in thin disks.  On the contrary, the well known MRI instability that is well suited to the geometry of a differentially rotating plasma cylinder is penalized by the thin disk geometry as this forces the instability driving factory to compete with a relatively strong bending of the magnetic filed lines.  Finally, the class of modes that we have analyzed can lead to the formation of a non-linear magnetic ``crystal'' structure, within the disk, made of a sequence of counter streaming toroidal current channels as described in Ref. [11] and may start the evolution of a classical disk configuration toward a sequence of plasma rings [4].

\acknowledgments
It is a pleasure to thank C. Crabtree, for his question on the role of the vertical component of the gravitational force that has stimulated the present analysis,  and L. Sugiyama for her timely comments. This work is sponsored in part by the U.S. Department of Energy.

\end{document}